\documentstyle[12pt,amssymbols]{article}
\input math_macros.tex

\begin{document}
\begin{titlepage}
\today          \hfill 
\begin{center}
\hfill    LBNL-38931 \\
\hfill    UCB-PTH-96/23 \\

\vskip .5in


\vskip .50in

{\large \bf 
  Analytic Study of Disoriented Chiral Condensates} 
\footnote{This work was supported in part by the Director, Office of 
Energy Research, Office of High Energy and Nuclear Physics, Division of 
High Energy Physics of the U.S. Department of Energy under Contract 
DE-AC03-76SF00098 and in part by the National Science Foundation under 
Grant PHY-95-14797.}

\vskip 1.0cm

Mahiko Suzuki

\vskip 0.5cm

{\em Department of Physics and Lawrence Berkeley National Laboratory\\
      University of California,
      Berkeley, California 94720}
\end{center}

\vskip .5in

\begin{abstract} 
   By introducing a quark source in the nonlinear $\sigma$ model, we obtain an
analytic boost-invariant solution as a candidate for the disoriented chiral
condensate (DCC) in $3 + 1$ dimensions.
  In order to trigger formation of the DCC, a strong transfer
of axial isospin charge must occur between the expanding source and the
interior in the baked Alaska scenario.  
An explicit chiral symmetry breaking is incorporated in the isospin-uniform
solution by connecting the decay period to the formation period. 
Quantitative estimates are presented with our simple solution.
At least in this class of solutions, the explicit symmetry breaking masks
almost completely the disorientation which would be reached asymptotically
in the symmetric limit.

\end{abstract}
\end{titlepage}
\renewcommand{\thepage}{\roman{page}}
\setcounter{page}{2}
\mbox{ }

\vskip 1in

\begin{center}
{\bf Disclaimer}
\end{center}

\vskip .2in

\begin{scriptsize}
\begin{quotation}
This document was prepared as an account of work sponsored by the United
States Government. While this document is believed to contain correct 
information, neither the United States Government nor any agency
thereof, nor The Regents of the University of California, nor any of their
employees, makes any warranty, express or implied, or assumes any legal
liability or responsibility for the accuracy, completeness, or usefulness%
of any information, apparatus, product, or process disclosed, or represents
that its use would not infringe privately owned rights.  Reference herein
to any specific commercial products process, or service by its trade name,
trademark, manufacturer, or otherwise, does not necessarily constitute or
imply its endorsement, recommendation, or favoring by the United States
Government or any agency thereof, or The Regents of the University of
California.  The views and opinions of authors expressed herein do not
necessarily state or reflect those of the United States Government or any
agency thereof, or The Regents of the University of California.
\end{quotation}
\end{scriptsize}

\vskip 2in
\begin{center}
\begin{small}
{\it Lawrence Berkeley Laboratory is an equal opportunity employer.}
\end{small}
\end{center}

\newpage

\renewcommand{\thepage}{\arabic{page}}
\setcounter{page}{1}
\section{Introduction}
     It has been speculated that disoriented chiral condensates (DCCs) may be
produced in high-energy hadron collisions and heavy-ion collisions
\cite{DCC}. In Bjorken's baked Alaska scenario \cite{BJ}, as
a hot dense matter spreads from a collision center, a disoriented vacuum
is created in its hollow interior. The disoriented vacuum eventually relaxes 
to the true vacuum by emitting the excess energy in soft pions. 
    Many numerical calculations have been performed with the $\sigma$ model to
study if a DCC can really be formed \cite{RW}. Though they often find a long 
range correlation leading to a DCC-like state, it is not long enough 
to lead to a spectacular Centauro or anti-Centauro event \cite{Centauro}.
Search for analytic solutions has also been done in the $\sigma$ model
 \cite{BlKr,AnBa,HuSu}. It shows among others that formation of isospin aligned
DCCs is strongly suppressed by the phase space of chiral rotations \cite{HuSu}. 

   Formation and decay of the DCC can be treated separately.
During the formation period, we may ignore the 
explicit chiral symmetry breaking due to the $u$ and $d$ current quark masses.
As the pion field cools down and its kinetic energy becomes comparable with
the energy scale of the explicit symmetry breaking, the approximation of 
chiral symmetry breaks down and thereafter the
symmetry breaking plays a major role.  The symmetry breaking causes 
attenuation of the classical pion field, and eventually quantum fluctuations 
dominate over the classical field. In other words, 
the classical field decays away by emitting pion quanta. 
 
In this paper we study analytically the formation and decay of the DCC by the 
nonlinear $\sigma$ model at zero temperature.  We assume that
there exists a window of time period, sometime after the initial stage 
of collision but before the beginning of decay, where the {\it chiral
symmetric} classical $\sigma$ model is a good approximation. Our main purpose
is to learn what initial condition triggers a DCC formation and 
how the DCC evolves subsequently.
We do not ask how a desired initial condition is created.  We will make 
one conceptual departure from our previous viewpoint \cite{HuSu} with regard 
to what we call the DCC in the $\sigma$ model. The DCC would not decay if
there were not for an explicit chiral symmetry breaking.  
For this reason, the terminal 
state of the pion field obtained in the chiral {\it symmetric} $\sigma$ model 
should be identified with the initial state of the decaying DCC.  Therefore the
DCC pion field should approach a nonvanishing asymptotic one at $t\rightarrow
\infty$ if a symmetry breaking is turned off.  

 In Sec.II we formulate our approach with the nonlinear $\sigma$ model coupled
to quarks. In Sec.III we consider several quark sources that
are of particular interest to the DCC formation in $1+1$ and $3+1$
dimensions. While a boost-invariant solution never reaches 
a static asymptotic configuration in $1+1$ dimensions, we find in $3+1$ 
dimensions a simple interesting source term leading to a boost-invariant pion  
field which approaches the true vacuum at $t\rightarrow\infty$.  
In Sec.IV, by making $SU(2)\times SU(2)$ rotations on this $3+1$ solution, 
we obtain the solutions in which the pion field 
approaches an isospin-uniform static configuration everywhere off the light 
cone. These solutions should be interpreted as the DCC. In Sec.V, we introduce 
an explicit symmetry breaking and obtain a complete spacetime evolution of 
the simple isospin-uniform solution of Sec.IV by smoothly continuing the 
formation period to the decay period. Because the ratio of the symmetry 
breaking quantity $m_{\pi}$ to the symmetric quantity $f_{\pi}$ is not small
numerically, the transition from the formation period to the decay period 
is obscure for most DCC solutions. In Sec.VI, we interpret our findings 
from the viewpoint of conservation of the axial isospin charge. It will help 
us to understand what can possibly lead to creation of the 
favorable quark sources and what is the chance to realize a favorable 
initial condition.

\section{Source of pion field}   

  In the environment of DCC formation, the matter particles 
are presumably in the quark-gluon phase rather than in the
nucleon phase. The appropriate low-energy effective Lagrangian is therefore the
$\sigma$ model coupled to quarks and antiquarks.  In terms of the 
chiral or current quark $q_{R,L}$ of $u$ and $d$, the Lagrangian in the 
$SU(2)\times SU(2)$ symmetry limit is:
\begin{eqnarray}  
 {\cal  L}& = & \frac{f_{\pi}^2}{4}{\rm tr}(\partial_{\mu}\Sigma^\dagger
              \partial^{\mu}\Sigma) 
               + i\overline{q}_R\!\not\!\partial q_R
               + i\overline{q}_L\!\not\!\partial q_L          \nonumber\\
      &  & \mbox{} - gf_{\pi}\,\overline{q}_L\Sigma q_R 
             - gf_{\pi}\,\overline{q}_R\Sigma^{\dagger} q_L,
\end{eqnarray}
where  
\begin{equation}
      \Sigma = e^{i{\mbox{\boldmath $\tau$}\cdot{\bf n}(x)\theta(x)}}.
\end{equation}
Three isospin components of pion field are identified with 
\begin{equation}
    \mbox{\boldmath $\pi$}(x) =f_{\pi}{\bf n}(x)\,\theta(x),
\end{equation}
where ${\bf n}(x)^2 = 1$.
We choose the nonlinear representation for the pion field instead of the linear
representation since we can impose more easily the condition that the $\pi-
\sigma$ fields be near the bottom of potential well.  By fixing the radial 
$\sigma$ field to $f_{\pi}$, we narrow the region of applicability of the 
$\sigma$ model to the energy range where the pion kinetic energy is much 
smaller than the depth of the {\it Mexican hat} potential:
\begin{equation}
 \frac{1}{2}\dot{\mbox{\boldmath $\pi$}}^2  
                              \ll  \frac{1}{8}m_{\sigma}^2 f_{\pi}^2.
\end{equation}
We move from the current quark $q_{R,L}$ to the constituent quark $Q_{R,L}$ by
\begin{equation}
          Q_R = \xi\, q_R, \,\,\,\,\,\, Q_L = \xi^{\dagger}q_L ,
\end{equation}
with
\begin{equation}
        \xi = e^{i{\mbox{\boldmath $\tau$}\cdot{\bf n}(x)\theta(x)/2}}.
\end{equation} 
Use of the constituent quark does not mean that the quarks are static in
the expanding shell. It is because the equation of motion takes a simpler form. 
The Lagrangian is now expressed in $\theta(x)$, ${\bf n}(x)$, and $Q_{R,L}(x)$ 
as
\begin{eqnarray}
 {\cal L}& = &\frac{f_{\pi}^2}{2}(\partial_{\mu}\theta\partial^{\mu}\theta +
            \sin^2\theta\,\partial_{\mu}{\bf n}\cdot\partial^{\mu}{\bf n}) +
               \frac{\lambda f_{\pi}^2}{2}({\bf n}^2 - 1)   \nonumber\\
         &   & \mbox{} +i\overline{Q}_R\!\not\!\partial Q_R 
              + i\overline{Q}_L\!\not\!\partial Q_L 
              - m_Q(\overline{Q}_L Q_R + \overline{Q}_R Q_L) \nonumber\\
   &   & \mbox{} + i\overline{Q}_R\xi\!\not\!\partial\xi^{\dagger} Q_R 
              + i\overline{Q}_L\xi^{\dagger}\!\not\!\partial\xi Q_L,
\end{eqnarray}
where $m_Q$ is the constituent quark mass given by $m_Q = gf_{\pi}$, and
$\lambda$ is a Lagrange multiplier.

  The Euler-Lagrange equation for $\theta$ is:
\begin{equation}
 \Box\theta - \sin\theta\,\cos\theta\,\partial_{\mu}{\bf n}\cdot\partial^{\mu}
             {\bf n} = - i\frac{m_Q}{f_{\pi}^2}{\bf n}\cdot(\overline{Q}
              \mbox{\boldmath $\tau$}\gamma_5 Q).
                   \label{theta}
\end{equation}
After $\lambda$ is eliminated, the Euler-Lagrange equation for ${\bf n}$ 
becomes
\begin{equation}
 \partial_{\mu}(\sin^2\theta\,{\bf n}\times\partial^{\mu}{\bf n})
            = -i\frac{m_Q}{f_{\pi}^2}{\bf n}\times(\overline{Q}
             \mbox{\boldmath $\tau$}\gamma_5 Q)\sin\theta.
                   \label{n} 
\end{equation}
We treat the quark field as a given external source.

    Through global $SU(2)\times SU(2)$ rotations a solution with a uniform 
isospin orientation (${\bf n}(x) = constant$) generates infinitely many
solutions with ${\bf n}(x)\neq constant$ that are degenerate in
energy \cite{AnBa, HuSu}.  It has been pointed out that in the boost-invariant 
$1+1$ dimensional case, the chiral rotations of isospin-uniform solutions 
generate all known solutions \cite{HuSu}. A wide class of isospin-nonuniform  
solutions can be obtained in this way in $3+1$ dimensions too. It is much 
easier to make chiral rotations of a uniform solution than to solve directly 
for general nonuniform solutions.  We study here the class of solutions that 
can be rotated into a uniform one by chiral rotations, {\it i.e.},
{\it Anselm-class} solutions according to the nomenclature of Bjorken.
The solutions of constant ${\bf n}$ are realized when the source points to 
a certain fixed direction everywhere:
\begin{equation}
  -i\biggl(\frac{m_Q}{f_{\pi}^2}\biggr)
      \overline{Q}\mbox{\boldmath $\tau$}\gamma_5 Q = {\bf n}_0\rho(x),
                     \label{j5}
\end{equation}
where $\rho(x)$ is a real Lorentz pseudoscalar function.
Once we set the isospin direction as in Eq.(\ref{j5}), the source term in the 
equation of motion for ${\bf n}$ disappears if we choose
\begin{equation}
           {\bf n}(x) = {\bf n}_0.
\end{equation}
Thus a configuration of uniform isospin orientation satisfies the
equation of motion for ${\bf n}(x)$ trivially.  
The equation of motion for $\theta(x)$ then turns into

\begin{equation}
     \Box \theta = \rho(x).
\end{equation}
This is the equation that we will focus on in Sec.III.  Since we have
already fixed the radial $\sigma$ field to its expectation value $f_{\pi}$, 
the scalar quark density $\overline{Q}Q$ does not contribute to formation of
the pion field. It is the isovector pseudoscalar density of the 
{\it constituent} $u$ and $d$ quarks that 
excites the phase pion field ${\bf n}_0\theta(x)$.

\vskip .2in       

\section{Quark densities of physical interest}

     We study the pion field generated by the typical quark sources which are
particularly interesting to the DCC formation.  We start with the case of
$1+1$ dimensions, which is viewed as the ${\bf p}_t = 0$ limit of the real 
world.

\vskip .2in

\noindent{\bf A. $1+1$ dimensions}

In order to solve Eq.(12), we need the Green function of $\Box\theta = 0$. 
The retarded Green function is 
\begin{equation}
     G(x;x') = \frac{1}{4}\Theta(t-t')\Bigl(\epsilon(t-t'-|z-z'|)
                        +\epsilon(t-t'+|z-z'|)\Bigr),
\end{equation}
where $\Theta(t)$ is the step function and $\epsilon(t)=\Theta(t) -
\Theta(-t)$. 
The simplest source is a pair of boost-invariant densities flying away
to opposite directions in the speed of light. This source is of practical
interest since a boost-invariant source in the coordinate space 
gives a boost-invariant pseudo-rapidity particle distribution. 
   With the variable $\tau' = \sqrt{t'^2 - z'^2}$, we express the source 
$\mbox{\boldmath$\rho$}(x') =
 -i(m_Q/f_{\pi}^2)(\overline{Q}\mbox{\boldmath $\tau$}\gamma_5 Q)$ as
\begin{equation}
    \mbox{\boldmath $\rho$}(x') = 
         {\bf q}_0\delta(\tau'^2)\Theta(t'), 
\end{equation}
where ${\bf q}_0$ is a constant vector in isospin space.
This form is the most general one that satisfies boost invariance, barring
derivatives of $\delta(\tau'^2)$. The source has a common isospin orientation 
at the both sides ($z'=t'$ and $z'=-t'$). It is singular at the origin $z'=t'=
0$ and dies away like $1/|z'|$ at far distances. Since the initial state at 
$t'=z'=0$ is outside the region of applicability of the $\sigma$ model, 
we will regularize $\mbox{\boldmath $\rho$}(x')$:
\begin{equation}
 {\bf n}_0\theta(x)  =  
             \int_{\varepsilon}^{t}dt'\Bigl(\int_{-\infty}^{-\varepsilon}dz'
              +\int_{\varepsilon}^{\infty}dz'\Bigr) G(x;x')
              \mbox{\boldmath$\rho$}(x'),
\end{equation}
where $x$ and $x'$ denote ($t, z$) and ($t', z'$), respectively.
Since the integral over $z'$ actually diverges if it is extended to
time $t' = 0 = z'$, a small neighborhood of 
$|z'|\leq \varepsilon$ has been excluded in the $z'$ integral.  The integrals 
over $z'$ at $\varepsilon <z'<\infty$ and $-\infty < z'<-\varepsilon$ 
are not separately boost invariant, but the sum is. The result is
\begin{equation}
  {\bf n}_0\theta(x) = \frac{1}{2}{\bf q}_0 
                 \ln\Biggl(\frac{\tau}{2\varepsilon}\Biggr),
\end{equation}
where $\tau = \sqrt{t^2 - z^2}$.
This logarithmic solution has been well known \cite{BlKr, HuSu}. It is  
peculiarity of $1+1$ dimensions that the boost-invariant pion field does 
not die away as $t \rightarrow \infty$ even at locations off the light cone.  
By global $SU(2)\times SU(2)$ rotations of this solution,
we can generate infinitely many more solutions, which are also known.

 It may be interesting to study for comparison the case opposite  
to boost invariance by
flipping the isospin direction for one side of the source pair, say, 
at $z'=-t'$. If the source splits statistically at random at $t=0$ 
the opposite isospin orientation will occur more frequently than the parallel 
orientation by isospin conservation. The solution for the opposite 
isospin orientation is not boost invariant: 
\begin{equation}
   {\bf n}_0\theta(x) = \frac{1}{4}{\bf q}_0
                   \ln\Biggl(\frac{t+z}{t-z}\Biggr).
\end{equation}

\noindent{\bf B. $3+1$ dimensions}

     The retarded Green function in $3+1$ dimensions is the well-known 
Li$\acute{e}$nard-Wiechert potential:
\begin{equation}
   G(x;x') = \frac{1}{2\pi}\delta\Bigl((t-t')^2 -
             ({\bf r}- {\bf r}')^2\Bigr)\Theta(t-t').
\end{equation}
The isovector source that flies away in a spherical shell in the speed 
of light can be expressed generally in the form
\begin{equation}
   \mbox{\boldmath $\rho$}(x') 
        = {\bf n}_0\sigma({\bf r'})\delta(\tau'^2)\Theta(t'),
\end{equation}
where $\tau'^2 = t'^2 - r'^2$. We first argue what $r'$-dependence is
physically interesting for $\sigma({\bf r'})$.

It appears natural that the total integrated strength of source 
dies away with time or distance as it feeds the pion
field. We will present an argument in favor of this behavior in the final
Section. Choice of $\sigma({\bf r}') = constant$ 
would make the source invariant under Lorentz boost along all directions. 
For such a source, however, the total source integrated over the shell would 
increase with time as $\sim t'$. It appears that more interesting possibility 
is
\begin{equation}
  \sigma({\bf r}')
   = \frac{\bar\sigma(\theta', \phi')}{r'^2}.
\end{equation}
For this $\sigma({\bf r}')$, 
the integrated source strength weakens like $1/t'$ or $1/r'$ 
since one power of $1/r'$ comes from $\delta(\tau'^2)$ in Eq.(19).
If the forward and backward patches of source within fixed solid angles
dominate in the formation of pion field, the source
behaves exactly like the boost-invariant $1+1$ dimensional source discussed
above.  For this source we obtain a remarkably simple pion field, 
\begin{equation}
 {\bf n}_0\theta(x) = {\bf n}_0\frac{\int\bar\sigma
                   (\theta', \phi')d\Omega'}{4\pi(t^2 - r^2)}. \label{solution}
\end{equation}
The pion field is determined only by the total source integrated over the shell,
independent of its distribution.  Furthermore, it is boost invariant.  
This extremely simple result is special to the $1/r'^2$ dependence 
of $\sigma({\bf r}')$.  
 
     If instead we demand that the source on a surface element within a solid
angle $\Delta\Omega$ should remain 
constant of time, the $r'$-dependence of the source is of the form
\begin{equation}
  \sigma({\bf r}') = \frac{\tilde\sigma(\theta', \phi')}{r'}.
\end{equation} 
In this case the produced pion field depends not only on the total strength of
source but also on the distribution on the shell. If $\tilde\sigma(\theta', 
\phi')$ is constant ($= \tilde\sigma_0$), the pion field is
\begin{equation}
   {\bf n}_0\theta(x) = \frac{1}{4r}{\bf n}_0\tilde\sigma_0
          \ln\Biggl(\frac{t+r}{t-r}\Biggr).
\end{equation}
If the isospin direction of source is opposite in sign in the forward and 
backward hemispheres ($\tilde\sigma_1$ = constant) as
\begin{equation}
   \tilde\sigma(\theta', \phi')  = \tilde\sigma_1 cos\theta',
\end{equation} 
the produced pion field is given by
\begin{equation}
   {\bf n}_0\theta(x) = 
          \frac{1}{4t}{\bf n}_0\tilde\sigma_1 
          \Biggl(\frac{t^2}{r^2} \ln\frac{t+r}{t-r} - \frac{2t}{r}\Biggr)
          \cos\theta
\end{equation}
We have considered here the class of $\sigma({\bf r}')$ that factorizes 
into $r'$ and ($\theta',\phi'$). For the sources of this kind, the relative
isospin strength on different parts of the source shell does not change with 
time. Since $t'= r'$ on the source, we can incorporate time 
variation of the relative isospin strength by adding nonleading $1/r'$ terms. 
However, it is the leading $1/r'$ term of $\sigma({\bf r}')$ that
determines the asymptotic pion field at $t \rightarrow \infty$. As for
isospin orientation, an aligned source $i\overline{Q}\mbox{\boldmath $\tau$}
\gamma_5 Q$ transforms nonlinearly under chiral rotations. Since the isospin 
direction is entangled with spacetime dependence in this transformation, 
isospin alignment of source is destroyed upon chiral rotations and the source 
becomes locally nonuniform. Instead of rotating the source, we can 
accomplish the same by rotating ${\bf n}_0\theta(x)$.
   The requirement that the total integrated pseudoscalar charge
$\int\mbox{\boldmath $\rho$}(x')d^3{\bf r}'$ should not increase with time sets
an upper bound on the $r'$ dependence of $\sigma(x')$:
\begin{equation}
     |\sigma(\bf r')|
   \leq \frac{|\sigma(\theta', \phi')|}{{\rm r'}}\;\;\;
    ({\rm r'}\rightarrow\infty).
\end{equation}
Then the asymptotic pion field at any location inside the light 
cone must vanish with this restriction:
\begin{equation}
    |\theta(x)| \leq \Biggl|\frac{1}{4\pi}\int\frac{\sigma(\theta',\phi')}
{t - r\cos\psi}d\Omega'\Biggr|\rightarrow 0 \;\;\;\;(t\rightarrow\infty),
          \label{asymp}
\end{equation}
where $\cos\psi = ({\bf r}\cdot{\bf r}')/rr'$.  We can see it explicitly in
the sample solutions given above.

   Do these isospin-uniform solutions describe DCCs in $3+1$ dimensions ?
One viewpoint is that any classical pion field should be called the DCC 
even if it is transient.  In this case the disorientation 
does not last forever even in the symmetric limit. We can take a different
viewpoint: The DCC is a disoriented state which would persist forever 
if the chiral symmetry were turned off. We have so far solved the 
$\sigma$ model in the chiral symmetry limit.  Therefore, if we take this second
viewpoint, what we should
identify with DCCs is not the solutions obtained above, but those solutions
that approach a nonvanishing limit, $\theta(\infty, {\bf r}) \neq 0$.
Our isospin-uniform solutions 
are driven to the true vacuum by chiral symmetric force 
alone. It turns out that the issue is largely semantic 
rather than physical. Later when we connect these solutions to the decay 
solutions by including an explicit symmetry breaking, we will find that 
the solutions with $\theta(\infty, {\bf r}) \neq 0$ oscillate
with a slightly larger amplitude during the decay period than those with
$\theta(\infty, {\bf r}) = 0$. Otherwise there is little difference between 
them. Though difference is very minor, there is an important conceptual 
distinction 
between the solutions with $\theta(\infty,{\bf r}) = 0$ and those with 
$\theta(\infty, {\bf r}) \neq 0$. In this paper we call the latter as the DCC.
In the next Section we will obtain the solutions with $\theta(\infty,{\bf r})
\neq 0$ by chiral rotations of the solutions with $\theta(\infty, {\bf r}) = 0$.

\vskip .2in
\section{DCC solutions}

   In this Section we obtain the DCC solutions 
with $\theta(\infty,{\bf r}) \neq 0$ from the
simple solution of Eq.(\ref{solution}) by chiral rotations.  Hereafter we
denote the solution ${\bf n}_0\theta(x)$ of Eq.(\ref{solution}) by 
${\bf n}_0\theta_0(x)$,
\begin{equation}
     \theta_0(x) = \frac{a}{t^2 - r^2} \;\;\;(\rightarrow 0\; {\rm as}\; t 
                   \rightarrow \infty),
\end{equation}
and the $\Sigma$ field of ${\bf n}_0\theta_0(x)$ by $\Sigma_0(x)$.
Let us designate global $SU(2)_R\times SU(2)_L$ rotations by a pair of vector 
rotation angles, $(2{\bf n}_R\theta_R, {2\bf n}_L\theta_L)$.  They transform
the nonlinear pion field $\Sigma_0(x)$ as
\begin{equation}
 \Sigma_0(x) \rightarrow 
     U({\bf n}_L\theta_L)\Sigma_0(x)\,U^{\dagger}({\bf n}_R\theta_R),
\end{equation}
where
\begin{equation}
   U({\bf n} \theta) 
   = e^{i\mbox{\boldmath $\tau$}\cdot{\bf n}\theta}.
\end{equation}
   It is straightforward to find the transformation formulas for 
$\theta(x)$ and ${\bf n}(x)$ \cite{Com}:
\begin{eqnarray}
  \cos\theta & = & \Bigl(c_L c_R + ({\bf n}_L\cdot{\bf n}_R)
                  s_L s_R\Bigr) c_0 + \nonumber\\
            &   & \Bigl(({\bf n}_0\cdot{\bf n}_R)c_L s_R -
                 ({\bf n}_0\cdot{\bf n}_L) s_L c_R + 
     ({\bf n}_0\times{\bf n}_L)\cdot{\bf n}_R\,s_L s_R\Bigr)s_0,
\end{eqnarray}
\begin{eqnarray}
{\bf n}\sin\theta & = &\Bigl({\bf n}_L s_L c_R - {\bf n}_R 
                       c_L s_R +({\bf n}_L\times{\bf n}_R)s_L
                       s_R \Bigr) c_0  \nonumber\\
                  &   & \mbox{} + \Bigl({\bf n}_0 c_L c_R +
                       ({\bf n}_0\times{\bf n}_L)s_L c_R
                       + ({\bf n}_0\times{\bf n}_R)c_L s_R \nonumber\\
               &   & \mbox{} +\bigl(({\bf n}_0\cdot{\bf n}_L){\bf n}_R +
           ({\bf n}_0\cdot{\bf n}_R){\bf n}_L - ({\bf n}_L\cdot
           {\bf n}_R){\bf n}_0\bigr)s_L s_R \Bigr)s_0,
\end{eqnarray}
where $c_{L,R}$ and $s_{L,R}$ stand for $\cos\theta_{L,R}$ and 
$\sin\theta_{L,R}$, respectively.  Since $\Sigma_0(x)$ approaches the unit 
matrix at $t=\infty$, the asymptotic pion field of the rotated solution is 
given by
\begin{equation}
      \Sigma(\infty, {\bf r}) 
       = U({\bf n}_L\theta_L)U^{\dagger}({\bf n}_R\theta_R).
\end{equation}
In terms of the $\theta$ and ${\bf n}$ fields,       
\begin{equation}
    \cos\theta(\infty, {\bf r}) = \cos\theta_L \cos\theta_R
              +  ({\bf n}_L\cdot{\bf n}_R)\sin\theta_L\sin\theta_R,
\end{equation}    
\begin{eqnarray}
{\bf n}(\infty, {\bf r})\sin\theta(\infty,{\bf r}) & = & {\bf n}_L
                               \sin\theta_L\cos\theta_R  \nonumber\\ 
                & & \mbox{} - {\bf n}_R\cos\theta_L\sin\theta_R 
                     + ({\bf n}_L\times{\bf n}_R)\sin\theta_L\sin\theta_R.
\end{eqnarray}
If we choose as a special case the rotations of ${\bf n}_R = {\bf n}_L$,
the asymptotic $\theta-{\bf n}$ fields take a very simple form:
\begin{equation}
      \theta(\infty, {\bf r}) = \theta_L - \theta_R, \;\;\;\;
      {\bf n}(\infty, {\bf r}) = {\bf n}_L .
\end{equation}
In this rotated solution, the pion field points 
asymptotically to the direction
of ${\bf n}_L (= {\bf n}_R)$ with magnitude $\theta_L - \theta_R$ everywhere
inside the light cone. At finite time, according to Eq.(32), the rotated 
${\bf n}(x)$ field is nonuniform in isospin direction.  
If furthermore ${\bf n}_L$ and 
${\bf n}_R$ are chosen along the directions of ${\bf n}_0$ of the 
solution ${\bf n}_0\theta_0(x)$, the rotated solution turns into an almost 
trivial form:
\begin{equation}
  {\bf n} \theta(x) = {\bf n}_0\Bigl(\theta_L - \theta_R + \theta_0(x)\Bigr).
\end{equation} 
This relation is a direct consequence of $SU(2)_R\times SU(2)_L$ symmetry
of Lagrangian.  When ${\bf n}_L$ (= ${\bf n}_R)$ does not coincide with 
${\bf n}_0$, we can still find relatively simple formulas that describe the 
asymptotic behavior of the nonuniform solution:
\begin{equation}
           \theta(x) \rightarrow \theta_L - \theta_R 
              +({\bf n}_L\cdot{\bf n}_0)\theta_0(x) + O(\theta_0(x)^2),
\end{equation} 
\begin{eqnarray}
    {\bf n}(x) & \rightarrow & {\bf n}_L +
      \frac{\sin(\theta_L +\theta_R)}{\sin(\theta_L-\theta_R)}
       ({\bf n}_0\times{\bf n}_L)\theta_0(x) \nonumber\\ 
       & & \mbox{} + \frac{\cos(\theta_L +\theta_R)}{\sin(\theta_L-\theta_R)}
       \Bigr({\bf n}_L\times({\bf n}_0\times{\bf n}_L)\Bigr)\theta_0(x) +
       O(\theta_0(x)^2),
\end{eqnarray}
where $\theta_0(x)$ is the asymptotic tail of the uniform solution
These formulas are valid for $\theta_0(x) \ll |\theta_L - \theta_R|$.

     In the hypothetical world of perfect chiral symmetry, the disoriented 
region of a DCC solution would expand without limit and no pions would be 
emitted. The massless pions would sit at rest forever since they cost no 
energy.  Actually the word "disoriented" is 
inappropriate in the symmetric limit because all chiral orientations 
are equivalent and related by symmetry.  In the real world the explicit
chiral symmetry breaking makes the solution of $\theta(\infty,{\bf r})\neq 0$
unstable by the amount $\Delta V =f_{\pi}^2 m_{\pi}^2(1-\cos\theta(\infty,
{\bf r}))$ in energy density.  By the time when the kinetic energy density of 
pion field decreases to be comparable to this energy density, 
our approximation of perfect chiral symmetry breaks down and we must start 
including an explicit chiral symmetry breaking. Let us make a quantitative 
estimate of this transition time.

    To be concrete, we consider the simple DCC solution of Eq.(37) with
$\theta_0(x)$ given by Eq.(28).
The asymptotic $\theta(x)$ in the absence of symmetry breaking is therefore 
\begin{equation}
  {\bf n}\,\theta(x) \rightarrow 
 {\bf n}_0\Bigl(\theta_L + \theta_R + \frac{a}{t^2 - r^2}\Bigr),\label{DCCsol} 
\end{equation}
where 
\begin{equation}
    a = \int\bar{\sigma}(\theta', \phi')d\Omega'/4\pi,
\end{equation}
according to Eq.(21). We can relate the magnitude of parameter $a$ to the time 
when the symmetry breaking becomes nonnegligible, namely the transition time
$t_0$ from the formation period to the decay period. Our chiral symmetric 
solutions do not make sense when the pion kinetic energy becomes comparable to
the potential energy of symmetry breaking; 
$\dot{\mbox{\boldmath $\pi$}}^2/2 \sim \Delta V$.  With our solution, 
this condition gives $2f_{\pi}^2 a^2/t^6 \sim m_{\pi}^2 f_{\pi}^2$ at locations 
away from the light cone. Therefore the symmetry breaking cannot be ignored 
after 
\begin{equation}
                t_0 = \biggl(\frac{|a|}{f_{\pi}}\biggr)^{1/3}.
\end{equation}
The transition time is delayed near the light cone where the pion field is
stronger than in the interior.  Since  
the spherical source expands nearly in the speed of light, $t_0$ is equal to
the radius of source $R_0$ at time $t_0$.  After time $t_0$, the DCC starts
decaying although the transition from formation to decay is not clear-cut.
Meantime the source keeps expanding and weakening in strength.
If $R_0$ is 5 {\it fm}, for instance, we obtain $|a| \sim 16 f_{\pi}^{-2}$. 
In order to generate a pion field of this magnitude and extent, we need, 
according to Eqs.(12), (19), (20), and (41), the integrated source strength of
\begin{equation}
\Bigl|\int i(\overline{Q}\mbox{\boldmath $\tau$}\gamma_5 Q)d^3{\bf r}\Bigr| 
    = \frac{2\pi f_{\pi}^2 a}{m_Q R_0}  \sim 13.
\end{equation}
The axial isospin charge density in the DCC can be 
computed with Eq.(40). It decreases with time at a fixed location and 
increases at a fixed time as we approach the source. At the origin 
it reaches a value independent of the source strength parameter $a$ 
by time $t_0$; $|{\bf A}_0(t_0,{\bf 0})| = 2f_{\pi}^{-3}$. 
We do not discuss how easily a source of this magnitude 
can be produced in hadron-hadron collisions.  We rather proceed 
to construct a continuous picture from formation to decay in the 
case of the simplest isospin-uniform DCC.
  
\vskip .2in
\section{Connecting formation to decay}

  In this Section we smoothly connect the simplest chiral symmetric solution 
of Eq.(\ref{DCCsol}) at $t < t_0$ to a solution at $t > t_0$ which attenuates 
with an explicit symmetry breaking. This boost-invariant solution is not only 
physically interesting but also easy to work with. Actually, as far as we have
explored, this is the only workable case where complete analytic study is
possible. We add the symmetry breaking 
\begin{equation}
  {\cal L}_{br} =  m_{\pi}^2 f_{\pi}^2 (\cos\theta - 1)
\end{equation}
to the chiral symmetric Lagrangian of Eq.(1) after time $t_0$, turning
the equation of motion for $\theta(x)$ into the sine-Gordon equation of $3+1$
dimensions. For the purpose of our analytic study, however, we reduce the 
equation of motion to a linear form by expanding $\cos\theta$ in 
${\cal L}_{br}$ around $\theta = 0$ and keeping only the leading term.  
This approximation breaks down near the light cone, where the phase pion field 
$\theta$ grows beyond $O(1)$, but it is good enough everywhere else for 
most purposes. Then the equation for $\theta$ is simply the equation of free 
motion with mass $m_{\pi}$. Since the chiral symmetric solution in the 
formation period is boost invariant, we look for a boost-invariant decay 
solution to connect to it. Inside the light cone where there is no quark 
matter, the $\theta$ field obeys the differential equation,
\begin{equation}
 \frac{1}{\tau^3}\frac{d}{d\tau}\biggl(\tau^3\frac{d}{d\tau}\theta\biggr)
       + m_{\pi}^2\theta = 0  \;\;\;\;\;\;\;(\tau^2 > 0).
\end{equation}
The general solution is given by the cylindrical functions:
\begin{equation}
  \theta(x) = \frac{c_1 J_1(m_{\pi}\tau)+c_2 N_1(m_{\pi}\tau)}{m_{\pi}\tau},
  \label{Bettersolution}
\end{equation}
where $J_1(z)$ and $N_1(z)$ are the Bessel and Neumann functions of the first 
order with $c_1$ and $c_2$ being constants. In the limit of 
$m_{\pi}\rightarrow 0$ ({\it i.e.,} $z \rightarrow 0$), 
$J_1(z)/z$ approaches $1/2$ while $N_1(z)/z$ behaves as
\begin{equation}
  \frac{N_1(z)}{z} \sim -\frac{1}{\pi}\Biggl(\frac{2}{z^2}-
      \ln\biggl(\frac{z}{2}\biggr) + \cdots  \Biggr), 
\end{equation}
We want the solution of Eq.(\ref{Bettersolution}) to turn into the chiral 
symmetric solution of Eq.(37) when $m_{\pi}\rightarrow 0$.  
This is realized if we choose the constants as
\begin{eqnarray}
     c_1 & = & 2(\theta_L - \theta_R) \equiv  2\theta_{\infty}, \nonumber\\
     c_2 & = & -\frac{1}{2}\pi m_{\pi}^2 a. 
\end{eqnarray}
At locations far away from the light cone, we find with the asymptotic 
formulas of the Bessel and Neumann functions that the pion field oscillates as
\begin{equation}
    \theta(x) \sim \sqrt{\frac{16\theta_{\infty}^2 +\pi^2m_{\pi}^4 a^2}
                    {2\pi(m_{\pi}\tau)^3}}
     \cos\Bigl(m_{\pi}\tau - \frac{3\pi}{4} +\vartheta_0\Bigr), 
      \;\;\;\;\;(\tau\rightarrow\infty)  \label{damp}
\end{equation} 
where
\begin{equation}
    \tan\vartheta_0 = \frac{\pi m_{\pi}^2 a}{4\theta_{\infty}}.
\end{equation}
Attenuation of $\theta(x)$ is interpreted as decay of the DCC due to emission 
of pions with nonvanishing mass. The decay is not exponential, 
but obeys a power law $\sim t^{-3/2}$ in amplitude with the time 
scale of $m_{\pi}^{-1}$.  The momentum spectrum of pions can be 
computed with the standard method \cite{HT}. 

Now that we have obtained a complete solution Eq.(\ref{Bettersolution}) with 
Eq.(48) from formation through decay including a symmetry breaking, 
it is appropriate here to make some quantitative discussion with our
solution.  Let us use for this purpose the set of parameters used earlier in 
Sec.IV; $t_0 = 5 fm/c$ and $|a| = 16 f_{\pi}^{-2}$.  For these values, the tail 
of the spacetime dependent term $a/(t^2-r^2)$ in the symmetric solution
at the transition time $t_0$ is comparable or larger than 
asymptotic limit $\theta_{\infty}$ ($|\theta_{\infty}| < \pi/2$).
As we go further down in time, the $a$-dependent part 
dominates over the $\theta_{\infty}$-dependent part by an order of magnitude 
in the asymptotic oscillation of Eq.(\ref{damp}). It means that 
distinction between the solution with $\theta_{\infty} = 0$ and those with 
$\theta_{\infty}\neq 0$ is insignificant even during the decay period. 
The origin of this rather unexpected result is traced back to the fact that the 
chiral symmetry breaking enters through $m_{\pi}$, which is numerically about 
the same in magnitude as the chiral symmetric energy scale $f_{\pi}$. 
If we set $m_{\pi}/f_{\pi}\rightarrow 0$ contrary to 
reality, the transition time $t_0$ would be stretched out as $t_0\sim
(f_{\pi}/m_{\pi})^{1/3}(af_{\pi}^2)^{1/3}f_{\pi}^{-1}\rightarrow \infty$. 
Then the tail of the $a$-dependent term of the pion field ($a/{t_0}^2 $) would
attenuate sufficiently by the time $t_0$ ($\theta(t_0) \sim 
(m_{\pi}/f_{\pi})^{2/3}(f_{\pi}^2 a)^{1/3}$) so that the $\theta_{\infty}$ 
term would dominate over the {\it a} term both at $t_0$ and in the asymptotic 
oscillation.  This does not happen at least in our solution. 
As for dependence on the strength of source $\rho(x)$, we can make the 
following statement: Since $a$ is proportional to the strength of source and 
$t_0$ depends on $a$ as $\sim a^{1/3}$, the transition 
is delayed for a stronger source, but not very sensitive to the 
source strength. Since the tail of the $a$-dependent term of $\theta(x)$ 
behaves like $\sim a^{1/3}$ at $t_0$, 
differentiating the DCC solution ($\theta_{\infty} \neq 0$) from the correctly 
oriented solution ($\theta_{\infty} = 0$) is a little harder when a DCC is 
generated by a stronger source. 
  
   The asymptotic $\theta(x)$ field of Eq.(\ref{damp}) is meaningless 
when the pion field becomes too weak.
This terminal time $t_f$, which can be interpreted as the
decay lifetime of DCC, is set by the condition $|\theta(t_f, {\bf r})| \ll 1$.  
According to Eq.(\ref{damp}), this happens at
\begin{equation}
        t_f \gg (2\pi f_{\pi}^4 a^2)^{\frac{1}{3}}m_{\pi}^{-1}.
\end{equation}
The right-hand side is $\sim 17 fm/c$ for $|a|=16f_{\pi}^{-2}$.  The decay 
proceeds very slowly in the time scale of hadron physics.

 In this Section, on the basis of the simplest isospin-uniform DCC, 
we have attempted to build a semiquantitative picture of 
formation and decay of DCC.  A missing information is how hadron-hadron 
collisions or heavy-ion collisions can possibly produce a strong enough
isovector pseudoscalar density of quark. We will develop some qualitative 
argument in the final Section.

\vskip .2in

\section{Discussion}

It is interesting to note that the equation of motion for $\theta(x)$ in Eq.(8)
is the statement of local conservation of the axial isospin current:
\begin{equation}
  {\bf n}(x)\cdot\partial^{\mu}{\bf A}_{\mu}(x) = 0.
\end{equation}
   When it is put into the form
\begin{eqnarray}
   {\bf n}\cdot\partial^{\mu}{\bf A}^{(\pi\sigma)}_{\mu} & = &
         \mbox{} - {\bf n}\cdot\partial^{\mu}{\bf A}^{(q)}_{\mu} \nonumber\\
        & = &  \mbox{} -im_Q
            {\bf n}\cdot(\overline{Q}\mbox{\boldmath $\tau$}\gamma_5 Q),
\end{eqnarray}
it is the equation of axial isovector charge transfer from the 
{\it current} quark to the $\mbox{\boldmath $\pi$}-\sigma$ system.  
The isovector pseudoscalar density of the {\it constituent} quark is the 
rate at which the transfer is made.

    The axial isospin charge ${\bf A}_0 = Q^{\dagger}(\mbox{\boldmath
$\tau$}/2)\gamma_5 Q$ takes a simple form in the infinite momentum limit.  
Let us examine for instance
the third component $A_0^{(3)}$. In the valence approximation, 
the axial isospin charge for the nucleon is simply related to $g_A=5/3$ 
of the famous SU(6) prediction or the constituent quark 
model prediction for the nuclear $\beta$-decay:  
\begin{eqnarray}
   \lefteqn{\lim_{{\bf p}\rightarrow\infty}
   \langle{\rm proton}({\bf p},h=\pm\frac{1}{2})|
    A_0^{(3)}|{\rm proton}({\bf p},h=\pm\frac{1}{2})\rangle}  \nonumber\\  
 & = & \lim_{{\bf p}\rightarrow\infty}\langle{\rm proton}({\bf p},
   h=\pm\frac{1}{2})| \frac{1}{|{\bf p}|}{\bf p}\cdot{\bf A}^{(3)}
     |{\rm proton}({\bf p}, h=\pm\frac{1}{2})\rangle   \nonumber\\
 & = & \pm \frac{1}{2} \times \frac{5}{3}, \nonumber\\ 
  \lefteqn{\lim_{{\bf p}\rightarrow\infty}
  \langle{\rm neutron}({\bf p},h=\pm\frac{1}{2})|
   A_0^{(3)}|{\rm neutron}({\bf p},h=\pm\frac{1}{2})\rangle}  \nonumber\\
 & = &\mbox{} -
  \lim_{{\bf p}\rightarrow\infty}\langle{\rm proton}({\bf p},h=\pm\frac{1}{2})|
   A_0^{(3)}|{\rm proton}({\bf p},h=\pm\frac{1}{2})\rangle.
\end{eqnarray}
For the quarks, $A_0^{(3)}$  
takes opposite signs for helicity $+1/2$ and $-1/2$ states and also for 
the $u$-quark and the $d$-quark. The sign remains the same for a quark and an
antiquark since ${\bf A}_{\mu}$ is even under charge conjugation.

   The axial isospin charge is small in the initial state of the $\bar{p}p$ 
collision.  Unless chiral symmetry is badly broken in hard collisions, the
total axial isospin charge of the quark-pion system should remain close to
this small number after collision.  In order to produce a large axial isospin
in the quark source on the shell, therefore, the small axial charge must  
be polarized into a pair of large charges of opposite signs, one in the 
quark-antiquark source within the shell and the other in the DCC pions.
Transfer of the axial charge between them acts as the source term to create
a DCC.  Our postulate on the time dependence of the quark 
source in Sec.III is consistent with this observation:
Increasing isovector pseudoscalar quark charge would mean 
that axial isospin charge transfer accelerates with time to polarize even 
further without limit.
A natural scenario is that a strong polarization of axial isospin
charge is created initially and gradually depolarizes as the 
source feeds the DCC pion field in the interior during the formation period.
 
However, in order for a constituent quark source to acquire a large axial 
isospin, the $u$-quark and the $d$-quark must react differently in the initial
hard collision:  According to Eq.(54), 
the helicity $+1/2$ state dominates over the $-1/2$ state for
the $u$-quark while the $-1/2$ state dominates over the $+1/2$ state for 
the $d$-quark. How can such a flavor dependence arise from the fundamental 
dynamics that is flavor blind ? A flavor dependence generated by a
statistically random process is far too small.
An answer to this question must come from quantum chromodynamics in high 
density and at high temperature. If creation of a large
isovector pseudoscalar density of quark is impossible, formation of a 
spectacular Centauro or an anti-Centauro would be ruled out.  
However, this does not prohibit formation of nonuniform DCCs which are
related to them through chiral rotations.  
This may be the reason why the classical pion field seen in the numerical
simulation \cite{RW} exhibits a textured isospin orientation, not a uniform
alignment of isospin.

\vskip .2in

\section{Acknowledgment}
   I have been benefited from discussions with J. D. Bjorken and from
   collaboration with Z. Huang at an early stage of work.  This work is
   supported in part by the Director, Office of Energy Research, Office of High
   Energy and Nuclear Physics, Division of High Energy Physics of the U.S.
   Department of Energy under Contract DE-AC03-76SF00098 and in part by the
   National Science Foundation under Grant PHY-90-21139.


\begin{thebibliography}{99}
\bibitem{Com} See also Reference 7. We have reproduced the formulas here since 
     there is one typographical error in those in Reference 7.
\end{thebibliography}
\end{document}